\documentclass[10pt,twocolumn]{article}

\usepackage[margin=0.72in]{geometry}
\usepackage{times}
\usepackage{graphicx}
\usepackage{booktabs}
\usepackage{multirow}
\usepackage{array}
\usepackage{tabularx}
\usepackage{enumitem}
\usepackage{caption}
\usepackage{subcaption}
\usepackage{amsmath}
\usepackage{xcolor}
\usepackage[hidelinks]{hyperref}
\usepackage{url}
\usepackage{microtype}
\usepackage{dblfloatfix}
\usepackage{float}
\usepackage{placeins}
\usepackage{listings}
\usepackage{balance}

\setlist[itemize]{leftmargin=*, topsep=2pt, itemsep=1pt, parsep=0pt}
\setlist[enumerate]{leftmargin=*, topsep=2pt, itemsep=1pt, parsep=0pt}

\renewenvironment{abstract}{%
    \vspace{-0.8em}
    \noindent{\bfseries Abstract}\par
    \vspace{0.35em}
    \small
    \noindent
}{%
    \par
    \vspace{0.6em}
    \normalsize
}

\newcommand{\keywords}[1]{%
    \noindent{\bfseries Keywords:} #1
    \par
    \vspace{0.8em}
}

\newcommand{\figureplaceholder}[2]{%
  \IfFileExists{#1}{%
    \includegraphics[width=#2]{#1}%
  }{%
    \fbox{\begin{minipage}[c][1.15in][c]{#2}
    \centering\small
    Figure placeholder\\[3pt]
    Add file: \texttt{#1}
    \end{minipage}}%
  }%
}

\title{\textbf{MindMirror: A Local-First Multimodal State-Aware Support System for Digital Workers}}

\author{%
Wenqi Luo$^{1}$ \quad Changbo Wang$^{1,*}$ \quad Yan Wang$^{1,*}$\\
$^{1}$School of Data Science and Engineering, East China Normal University\\
\texttt{luowenqi52@gmail.com}\\
$^{*}$Corresponding authors: \texttt{cbwang@dase.ecnu.edu.cn}, \texttt{yanwang@dase.ecnu.edu.cn}
}

\date{}

\begin{document}

\maketitle
\vspace{-1.0em}

\begin{abstract}
Digital workers often experience fatigue, anxiety, reduced attention, and task blockage during prolonged computer-based work. Existing productivity tools mainly focus on task completion and output management, while general-purpose AI chatbots typically require users to actively formulate their problems before receiving useful assistance. This paper presents MindMirror, a local-first multimodal state-aware support system for digital workers. MindMirror integrates camera-based facial expression cues, text input, optional speech interaction, structured blockage reflection, local large language model (LLM)-based response generation, and daily/weekly review reports. The system forms a closed workflow of state checking, manual correction, structured articulation, suggestion generation, and state review. The current prototype follows a local-first design, while optional speech services may rely on third-party APIs when enabled. The prototype is implemented with a Web frontend, Flask backend, an emotion recognition model, an Ollama-hosted Qwen model, Chart.js visualization, and local JSON/LocalStorage session records. For evaluation, we assess the emotion recognition module on an independent seven-class image-level facial expression benchmark containing 6,767 images. The fine-tuned Hugging Face model improves accuracy from 59.66\% to 94.49\% compared with the non-fine-tuned checkpoint baseline, corresponding to an absolute improvement of 34.83 percentage points on this benchmark. We further validate the prototype through endpoint-level reliability tests, voice-interaction latency tests, and a small formative user feedback study with six digital workers. The formative results suggest that users value the local-first design, manual correction mechanism, and structured reflection workflow. We discuss system boundaries, artifact availability, privacy considerations, and limitations, emphasizing that MindMirror is not intended for psychological diagnosis but serves as a lightweight, user-controllable tool for state reflection and supportive interaction.
\end{abstract}


\section{Introduction}

With the growth of remote work, software development, academic writing, design, and other computer-based activities, many users spend long periods in intensive digital work environments. Digital workers not only need to complete tasks but also need to manage attention loss, emotional fluctuations, cognitive overload, and moments of being ``stuck'' without being able to clearly explain why. Workplace mental health has also become an important global issue: the World Health Organization reports that depression and anxiety are associated with substantial productivity losses worldwide \cite{who_work_mental_health}.

Existing tools remain insufficient in such scenarios. Traditional productivity tools often emphasize task decomposition, time management, and output tracking, which may create additional pressure to continue working but provide limited support for understanding the user's current state. General-purpose AI chatbots can generate flexible responses, but they typically require users to first formulate a clear request. When users are tired, anxious, or blocked, formulating the problem itself can become a burden.

To address this gap, we present \textit{MindMirror}, a local-first multimodal state-aware support system for digital workers. The goal of MindMirror is not to diagnose mental disorders or replace professional psychological services. Instead, it provides low-friction state recording, structured articulation, and lightweight support in everyday work contexts. MindMirror uses camera images, text input, and optional speech interaction to support state awareness, while always allowing users to manually correct the emotion recognition result.

The system applies a three-question structured reflection workflow, asking users where they are stuck, what they have tried, and what they want to achieve next. A local LLM generates a three-step supportive suggestion, and daily/weekly reports help users review state changes over time.

The main innovations of this paper are organized as follows:
\begin{itemize}
    \item \textbf{Affective computing for digital workers.} We formulate affective computing for digital workers as a state-reflection and blockage-support problem, rather than as a productivity-optimization or clinical-diagnosis problem. MindMirror treats fatigue, anxiety, attention reduction, and work blockage as everyday interaction states that can be recorded, corrected, reflected on, and reviewed during digital work.
    \item \textbf{Multimodal human-computer interaction.} We design a multimodal human-computer interaction workflow that combines camera-based facial expression cues, text input, optional speech interaction, manual emotion correction, structured blockage reflection, local LLM-based suggestion generation, and visualized daily/weekly review. This workflow reduces the burden of open-ended prompt formulation and keeps the user in control of state interpretation.
    \item \textbf{System implementation and evaluation.} We develop a working MindMirror prototype with a Web frontend, Flask backend, emotion recognition model, Ollama-hosted Qwen model, Chart.js visualization, and local JSON/LocalStorage records. We evaluate the prototype through emotion recognition benchmarking, system-level technical validation, and formative user feedback.
\end{itemize}

\section{Related Work}

\subsection{Affective Computing and Emotion Sensing}

Affective computing studies computational systems that can recognize, interpret, and respond to human affective states. Picard's foundational work introduced affective computing as a framework for designing computers that can sense and respond to human emotions \cite{picard_affective}. Later surveys on affect detection further showed that affective states can be inferred from multiple channels, including facial expressions, speech, physiology, text, body language, and multimodal behavioral signals \cite{calvo_affect_detection}. These studies provide the broader foundation for MindMirror's multimodal state-aware design.

Unlike systems that treat affect recognition as an automatic final judgment, MindMirror treats emotion recognition as an initial state cue. This design choice is important because affective states are subjective, context-dependent, and often difficult to infer reliably from a single signal. Therefore, the system allows users to manually confirm or correct the detected state before it is stored or used in reflection.

\subsection{Facial Expression Recognition}

Facial expression recognition (FER) has been widely studied as a visual cue for affective computing and human-computer interaction. The FER-2013 challenge introduced a widely used seven-class facial expression benchmark \cite{goodfellow_fer2013}, and CK+ remains a classic dataset for action-unit and emotion-specified facial expression analysis \cite{lucey_ckplus}. Recent deep FER surveys highlight that in-the-wild FER remains challenging due to illumination, pose, occlusion, identity bias, limited data, and subjective labels \cite{li_deepfer_survey}.

Transformer-based architectures have also been explored for FER. For example, TransFER uses transformers to learn relation-aware facial expression representations \cite{xue_transfer}, while mask-based vision transformer models further explore transformer architectures for in-the-wild FER \cite{li_mvt}. These studies motivate our use of a ViT-based emotion recognition model. In MindMirror, however, the predicted facial expression label is not interpreted as a definitive psychological state. Instead, it is used as an editable image-level cue within a user-controllable workflow.

\subsection{Multimodal Emotion Recognition}

Multimodal emotion recognition combines signals such as face, speech, and text to better understand human affective states. Recent surveys define multimodal emotion recognition as the integration of text, speech, face, and other modalities, and emphasize its relevance to HCI \cite{lian_mer_survey}. Multimodal datasets such as CMU-MOSEI further show the value of modeling interactions among language, audio, and visual modalities for emotion and sentiment analysis \cite{zadeh_mosei}.

MindMirror follows this multimodal direction but remains conservative in how it uses inferred affective cues. The current prototype supports visual state checking, text input, optional speech interaction, and manual state selection. The goal is not to produce a fully automated diagnosis, but to combine multimodal cues with user confirmation and reflection.

\subsection{LLM-Based Supportive Interaction}

Large language models have created new opportunities for supportive dialogue and mental health-related applications. Recent systematic reviews show that LLMs have been explored for early screening, digital interventions, conversational agents, and clinical support, while also emphasizing risks related to reliability, safety, evaluation, privacy, and over-reliance \cite{guo_llm_mental}. Earlier conversational agents such as Woebot demonstrated that automated dialogue systems can be feasible and acceptable for self-help contexts, but such systems require careful boundary-setting and evaluation \cite{fitzpatrick_woebot}.

MindMirror adopts LLMs for lightweight supportive interaction rather than clinical intervention. The system first guides users through structured blockage reflection and then generates a concise three-step suggestion. This design reduces open-ended prompt formulation burden while avoiding claims of diagnosis or therapy.

\subsection{Reflection, Human-AI Interaction, and Privacy}

MindMirror's review module is related to personal informatics and reflective technologies. Li, Dey, and Forlizzi proposed a stage-based model of personal informatics consisting of preparation, collection, integration, reflection, and action \cite{li_personal_informatics}. Reflective informatics further argues that computational systems can be designed to support reflection in different ways \cite{baumer_reflective}.

Human-AI interaction guidelines emphasize that AI systems should support appropriate user expectations, allow efficient correction, and make system uncertainty manageable \cite{amershi_guidelines}. MindMirror follows this principle by allowing users to manually correct emotion cues and by treating model output as an editable suggestion. This is especially important for affective computing because emotion recognition technologies raise ethical concerns around surveillance, autonomy, accuracy, and sensitive emotional data. MindMirror therefore adopts a local-first design. Local-first software argues that users should own their data despite the convenience of cloud services \cite{kleppmann_localfirst}. In our system, state records and review data are primarily stored locally, and temporary media files should be deleted after processing.

\section{Design Goals}

MindMirror is guided by four design goals.

\textbf{DG1: Low-friction state checking.} The system should allow users to quickly record their current state without interrupting their workflow. A state check can be initiated through camera-based cues or manual selection.

\textbf{DG2: User-controllable emotion recognition.} Facial expression recognition may be affected by lighting, camera angle, occlusion, and individual differences. Therefore, model output should not be treated as an authoritative judgment. Users must be able to confirm, modify, or ignore the recognized result.

\textbf{DG3: Structured articulation before suggestion generation.} Instead of presenting a blank chat box, the system asks three questions: where the user is stuck, what they have tried, and what they want to achieve next.

\textbf{DG4: Local-first state review.} The system should not only provide immediate responses but also store user-confirmed states and reflection records locally, generating daily/weekly reports and trend visualizations to support longer-term self-reflection.

\section{System Architecture and Implementation}

\subsection{Overall Architecture}

MindMirror adopts a five-layer architecture: user interaction layer, frontend layer, backend service layer, AI engine layer, and data storage layer. The user interaction layer includes a Web application, local workstation entry, and an optional VSCode extension. The frontend layer uses HTML5, CSS3, JavaScript, the MediaDevices API, and Chart.js. The backend service layer is implemented with Flask and provides APIs for health checking, emotion analysis, voice chat, and audio file service. The AI engine layer includes an emotion recognition model, an Ollama-hosted Qwen model, and speech STT/TTS. The data storage layer includes browser LocalStorage and local JSON session records.

The main entry point is a local Web application. Users can view today's summary on the homepage, perform emotion recognition or manual confirmation in the state-check page, complete structured input in the blockage-reflection page, view state trends and suggestion summaries in the review-report page, and inspect or delete past records in the history page.

Representative screenshots of the current prototype are shown in Figures~\ref{fig:ui-dashboard-state} and~\ref{fig:ui-reflection-review}. The screenshots use the current Chinese interface implementation; the interaction workflow itself is language-independent and can be localized without changing the system design.

\begin{figure}[t]
\centering
\figureplaceholder{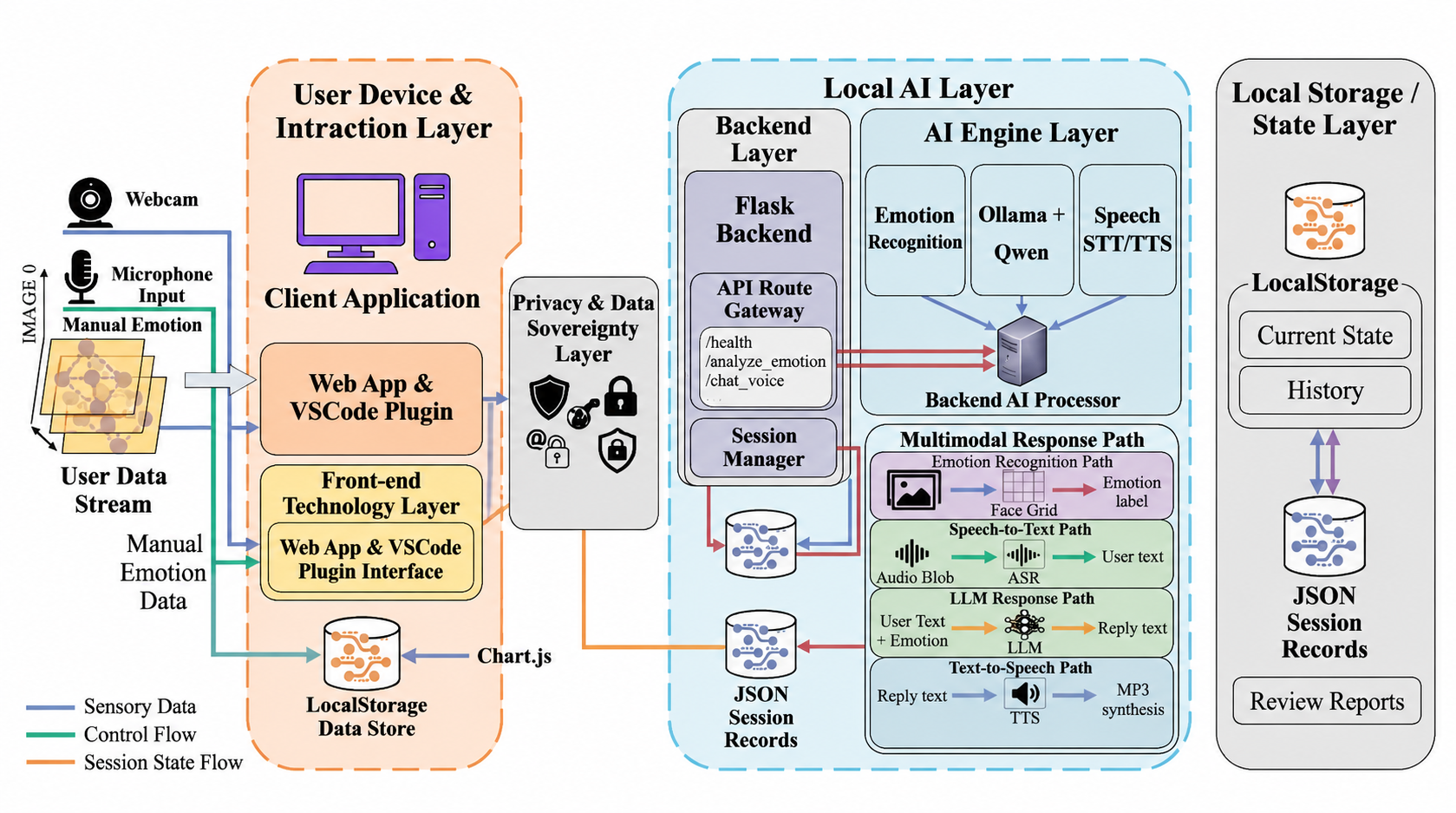}{\columnwidth}
\caption{Overall architecture of MindMirror. The system connects user-side interaction, a Web frontend, Flask backend/API routing, local AI engines, and local session storage.}
\label{fig:architecture}
\end{figure}

\subsection{State Check and Manual Correction}

The state-check module accesses the camera through the browser MediaDevices API. After the user starts the camera, the frontend captures a video frame with Canvas and converts it into a base64 image, which is sent to the backend emotion-analysis endpoint. The backend performs emotion recognition and returns an emotion label.

At the interaction level, MindMirror does not require users to accept the model judgment. The predicted label is only a default suggestion, and users can manually select or correct their current state. This mechanism reduces the risk of model misclassification and mitigates the discomfort of being monitored or judged.

\subsection{Structured Blockage Reflection}

Traditional AI chat interfaces often start with a blank input box, but users may not know how to begin when tired or blocked. MindMirror therefore introduces a structured reflection workflow that asks three questions:
\begin{enumerate}
    \item Where am I stuck?
    \item What have I tried?
    \item What do I want to achieve next?
\end{enumerate}

The system combines the user-confirmed state, the three reflection fields, and recent session context into a prompt for the local LLM. The generated response is a three-step supportive suggestion. The goal is not to provide medical advice, but to help users convert vague stress into small, actionable steps.

\subsection{Speech and Text Input}

Text input is the primary interaction path for structured reflection. MindMirror also implements optional speech interaction: the frontend records WebM audio and submits it to the voice-chat endpoint. The backend converts the audio format, invokes speech-to-text when a speech API is configured, generates a response using the LLM, and can synthesize the response into speech.

Because speech interaction is an auxiliary convenience path rather than the core research contribution, this paper validates the voice-chat path with a small fixed-audio test set rather than treating speech recognition as a primary research target. The current speech pipeline may rely on third-party speech APIs, so the system is described as local-first rather than fully offline. Future versions can replace this component with local ASR/TTS models to reduce external dependencies.

\subsection{Review Reports and History}

MindMirror records user-confirmed states, blockage inputs, AI suggestions, and timestamps locally for daily/weekly reports and history review. A review report includes state distribution, state trends, typical blockage summaries, suggestion summaries, and next-step reminders. The frontend uses Chart.js to display trend charts and state distribution charts.

This design makes MindMirror more than a one-time chatbot. It becomes a long-term state reflection tool. Users can review how their state changes across different work periods and delete records whenever needed.

\section{Interaction Workflow and Prompt Design}

\subsection{End-to-End Workflow}

\begin{figure*}[t]
\centering
\figureplaceholder{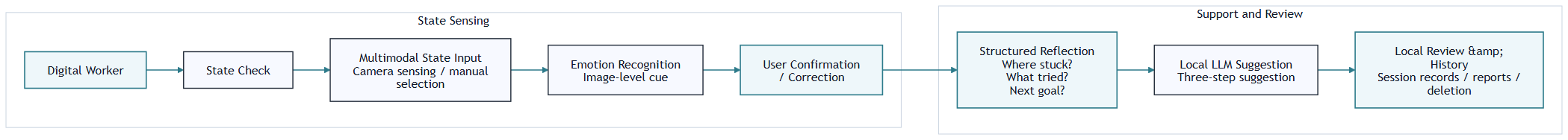}{0.92\textwidth}
\caption{End-to-end workflow of MindMirror. The system supports local-first multimodal state checking, user confirmation or correction of emotion cues, structured reflection, local LLM-based suggestion generation, and local review/history management.}
\label{fig:workflow}
\end{figure*}

To clarify how MindMirror forms a complete interaction loop rather than a set of isolated functions, we summarize the end-to-end workflow in Figure~\ref{fig:workflow}. The workflow consists of three stages: state checking, structured blockage reflection, and local review.

The first stage is state checking. Users can record their current state through either camera-based input or manual input. When the camera-based path is used, the frontend obtains the video stream through the MediaDevices API, captures a frame with Canvas, converts it into a base64 image, and sends it to the backend emotion analysis endpoint. The emotion recognition model then produces an image-level facial expression label. This label is not treated as the final state. Instead, it is displayed as an initial state cue, and the user can either confirm or manually correct it.

The second stage is structured blockage reflection. Rather than presenting an open-ended chat box, MindMirror asks three questions: where the user is stuck, what they have tried, and what they want to achieve next. The confirmed state label, the structured reflection content, and recent session history are assembled into a local prompt and sent to the local LLM. The model then generates a concise three-step supportive suggestion. This design reduces the burden of prompt formulation when users are already tired, stressed, or cognitively blocked.

The third stage is local review. The confirmed state, blockage description, generated suggestion, and timestamp are stored as local session records through LocalStorage or JSON files. These records are used to generate daily/weekly review reports, including state trends, typical blockage summaries, suggestion summaries, and next-step reminders. Users can review or delete historical records at any time.

\begin{figure*}[t]
\centering
\figureplaceholder{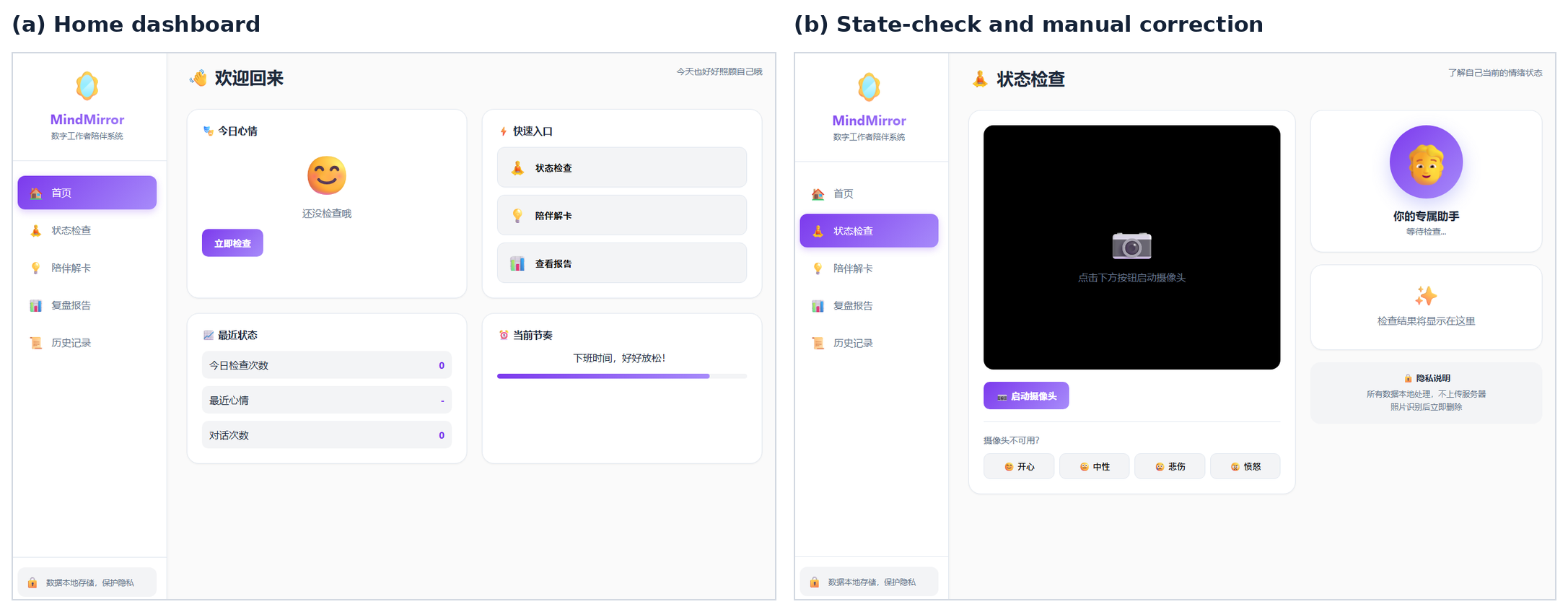}{0.96\textwidth}
\caption{Representative user interfaces of MindMirror: (a) the homepage dashboard and quick-entry panel, and (b) the state-check page with camera input, assistant status, recognition result area, privacy notice, and manual state options.}
\label{fig:ui-dashboard-state}
\end{figure*}

\begin{figure*}[t]
\centering
\figureplaceholder{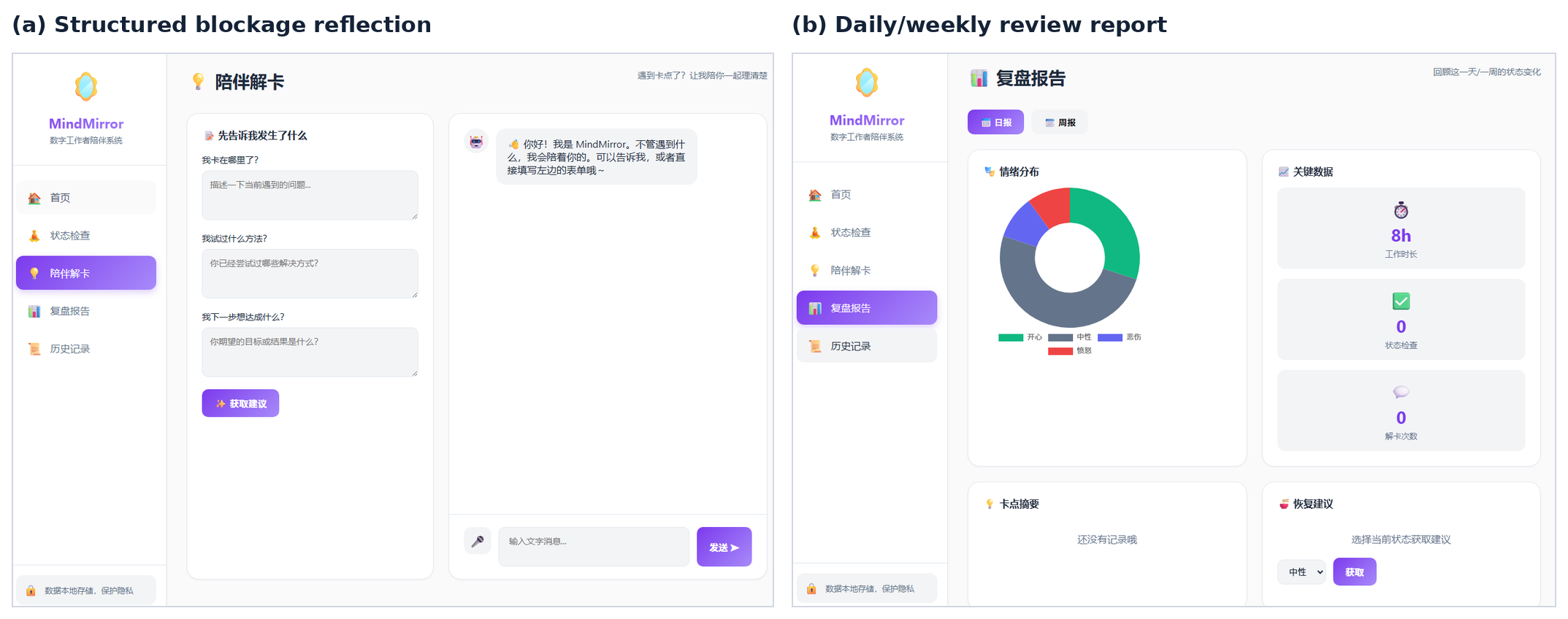}{0.96\textwidth}
\caption{Representative interaction screens after state checking: (a) the structured blockage-reflection page with three guiding questions and chat support, and (b) the review-report page with state distribution, key statistics, blockage summary, and recovery suggestion areas.}
\label{fig:ui-reflection-review}
\end{figure*}

Figures~\ref{fig:ui-dashboard-state} and~\ref{fig:ui-reflection-review} complement the abstract workflow by showing how the main interaction stages appear in the current Web prototype. The homepage and state-check page lower the entry barrier to state recording, while the reflection and review pages make the blockage-support loop visible and revisitable.

\subsection{Prompt Template}

The prompt is designed to make the LLM response structured, bounded, and non-clinical. Table~\ref{tab:prompt-fields} summarizes the main fields passed to the LLM. The system instruction asks the model to provide supportive three-step suggestions, while explicitly avoiding medical diagnosis or treatment recommendations.

\begin{table}[H]
\centering
\caption{Input fields used in the MindMirror prompt.}
\label{tab:prompt-fields}
\scriptsize
\setlength{\tabcolsep}{3pt}
\begin{tabularx}{\columnwidth}{p{0.36\columnwidth}X}
\toprule
Field & Description \\
\midrule
detected\_emotion & Initial facial-expression cue predicted by the model. \\
user\_confirmed\_state & User-confirmed or manually corrected state label. \\
reflection\_goal & What the user wants to achieve next. \\
reflection\_blockage & Where the user reports being stuck. \\
reflection\_tried & What the user has already tried. \\
session\_context & Recent local session context, when available. \\
\bottomrule
\end{tabularx}
\end{table}

The output format is a three-step suggestion: an immediate action, a short-term strategy, and a longer-term reminder. Each step contains a concise action and a short explanation. The prompt also includes safety constraints: the model should not diagnose, should not provide medical or therapeutic treatment, should use supportive language, and should recommend professional help when the user expresses severe or persistent distress. Appendix~\ref{appendix:prompt-study} provides the template and questionnaire items used in the formative study.

\subsection{Safety Boundary}

MindMirror uses affective cues only as editable state references. The LLM is instructed to generate general support for work-related blockage rather than clinical interpretation. A typical safety statement is: ``This system provides general emotional support and productivity-oriented suggestions, but it cannot replace professional psychological counseling or medical services. If you experience persistent or severe distress, please seek professional help.''

\section{Technical Report Scope and Artifact Availability}

This manuscript is prepared as a technical report. The goal is to document the current MindMirror prototype, implementation choices, evaluation logs, and design limitations. The report should be read as a system and HCI contribution rather than a clinical evaluation or a new machine-learning algorithm paper.

\begin{table}[H]
\centering
\caption{Technical report scope and artifact availability.}
\label{tab:artifact}
\scriptsize
\setlength{\tabcolsep}{3pt}
\begin{tabularx}{\columnwidth}{>{\raggedright\arraybackslash}p{0.30\columnwidth}>{\raggedright\arraybackslash}X}
\toprule
Item & Description \\
\midrule
Repository & \href{https://github.com/lwq-NEWCEO/mindmirror}{github.com/lwq-NEWCEO/mindmirror} \\
License & MIT License. \\
Included code & Web frontend, Flask backend, emotion-recognition integration, LLM interaction module, and visualization components. \\
Public data & FER-2013 can be obtained from public dataset sources. \\
Self-collected data & 1,354 supplementary Asian facial-expression images may be provided upon reasonable request, subject to privacy and consent constraints. \\
Model checkpoint & Fine-tuned from the Hugging Face \texttt{vit-FER} checkpoint. \\
Reproducibility & Training scripts, configuration files, and runtime instructions are intended to support reproduction of the prototype. \\
\bottomrule
\end{tabularx}
\end{table}

\begin{table}[H]
\centering
\caption{Runtime environment used by the prototype.}
\label{tab:runtime}
\scriptsize
\setlength{\tabcolsep}{2.5pt}
\begin{tabularx}{\columnwidth}{p{0.25\columnwidth}p{0.25\columnwidth}X}
\toprule
Component & Version & Role \\
\midrule
Python & 3.10+ & Backend environment \\
Flask & 2.3+ & Web service framework \\
Ollama & 0.1.30+ & Local LLM runtime \\
Qwen & 7B/14B & Local language model \\
PyTorch & 2.0+ & Deep-learning framework \\
Transformers & 4.30+ & Hugging Face model interface \\
Chart.js & 4.4+ & Frontend visualization \\
GPU & RTX 3090/4090 recommended & 12GB+ VRAM recommended for local model use \\
\bottomrule
\end{tabularx}
\end{table}

\section{Evaluation}

We evaluate MindMirror from three perspectives: emotion recognition benchmarking, system-level technical validation, and formative user feedback. The purpose of the evaluation is to validate the feasibility of the prototype and its interaction workflow, not to claim clinical efficacy or psychological diagnosis.

\subsection{Emotion Recognition Benchmark Setup}

We compare a non-fine-tuned baseline with a fine-tuned model on an independent seven-class image-level facial expression benchmark containing 6,767 labeled images. The benchmark covers angry, disgust, fear, happy, neutral, sad, and surprise. Under the reporting protocol used in this technical report, this 6,767-image benchmark is treated as the denominator of the headline accuracy comparison. It is not used for model training, validation, hyperparameter selection, or early stopping.

The baseline is a non-fine-tuned checkpoint baseline: the publicly available ViT-based facial expression recognition checkpoint is directly used for inference without task-specific fine-tuning on the target training data. The fine-tuned model uses the same ViT-based checkpoint as the backbone and is further optimized on the target training split. The input resolution is $224\times224\times3$, the batch size is 32, the optimizer is AdamW, and the learning rate is $5\times10^{-5}$. Training was stopped according to validation loss.

To reduce the risk of data leakage, duplicate and low-quality samples were removed during preprocessing. We also performed train/evaluation separation, manual inspection, and image-similarity checking. Following general machine-learning evaluation practice, the benchmark used for reporting final accuracy is kept separate from model selection to avoid inflated performance \cite{kaufman_leakage}.

\begin{table}[H]
\centering
\caption{Main emotion-recognition benchmark and model summary.}
\label{tab:benchmark-summary}
\scriptsize
\setlength{\tabcolsep}{2.5pt}
\begin{tabularx}{\columnwidth}{p{0.31\columnwidth}X}
\toprule
Item & Description \\
\midrule
Evaluation benchmark & Independent seven-class facial-expression benchmark, 6,767 images \\
Classes & angry, disgust, fear, happy, neutral, sad, surprise \\
Source composition & FER-2013 images plus self-collected supplementary Asian facial-expression images \\
Preprocessing & low-resolution, severely blurred, occluded, and inconsistent-label samples removed \\
Baseline & Non-fine-tuned Hugging Face ViT checkpoint \\
Fine-tuned model & Same ViT checkpoint, fine-tuned on the target training split \\
Training configuration & $224\times224\times3$ input, batch size 32, AdamW, learning rate $5\times10^{-5}$, early stopping by validation loss \\
Leakage prevention & duplicate removal, evaluation separation, manual inspection, and image-similarity checking \\
Reported scope & image-level facial expression classification; not psychological state recognition \\
\bottomrule
\end{tabularx}
\end{table}

\subsection{Overall Accuracy}

The baseline model correctly classifies 4,037 out of 6,767 benchmark images, reaching an accuracy of 59.66\%. The fine-tuned model correctly classifies 6,394 out of 6,767 images, reaching an accuracy of 94.49\%. This corresponds to an absolute improvement of 34.83 percentage points.

Because the benchmark focuses on image-level facial expression classification, this result should be interpreted as evidence that fine-tuning improves the stability of the visual state cue used by MindMirror, not as evidence of real psychological state recognition. In the deployed workflow, the model output is only an initial and editable cue. The user can confirm, correct, or ignore the result before it is stored or used in structured reflection.

\begin{table}[H]
\centering
\caption{Overall benchmark accuracy comparison.}
\label{tab:overall-accuracy}
\scriptsize
\begin{tabular}{lccc}
\toprule
Model & Correct & Total & Accuracy \\
\midrule
Baseline model & 4,037 & 6,767 & 59.66\% \\
Fine-tuned model & 6,394 & 6,767 & 94.49\% \\
\bottomrule
\end{tabular}
\end{table}

\subsection{Per-Class Accuracy Analysis}

Class-level inspection shows that the baseline model is relatively stronger on visually distinctive expressions such as happy, but weaker on several negative or ambiguous expressions such as sad, fear, and angry. After fine-tuning, performance improves across all seven expression categories in the evaluation log. These results suggest that task-specific adaptation can make the image-level facial-expression signal more useful as an initial interaction cue.

However, class-level results should still be interpreted conservatively. Static facial images cannot reliably reveal a user's actual psychological state, and visual expression categories may overlap because of lighting, pose, occlusion, identity differences, and subjective labeling. Therefore, MindMirror keeps manual correction as a necessary part of the interaction design rather than treating any class prediction as a final judgment. Detailed per-class accuracy results are provided in Appendix~\ref{appendix:perclass}.

\subsection{Diagnostic Subset Metrics}

In addition to the main 6,767-image benchmark, we report detailed metrics on a 500-sample diagnostic subset for which prediction vectors are available. This subset focuses on three frequently confused or practically important classes: disgust, fear, and surprise. The model correctly classifies 448 out of 500 samples, yielding an accuracy of 89.6\%. Because this subset does not cover all seven classes as ground-truth categories, it should be interpreted as a diagnostic analysis rather than a replacement for the full benchmark.

\begin{table}[H]
\centering
\caption{Precision, recall, and F1-score on the diagnostic subset.}
\label{tab:diagnostic-prf}
\scriptsize
\setlength{\tabcolsep}{4pt}
\begin{tabular}{lcccc}
\toprule
Emotion & Precision & Recall & F1-score & Support \\
\midrule
disgust & 0.9915 & 0.8478 & 0.9141 & 138 \\
fear & 0.9398 & 0.8741 & 0.9058 & 143 \\
surprise & 0.9364 & 0.9406 & 0.9385 & 219 \\
Macro avg. & 0.9559 & 0.8875 & 0.9195 & 500 \\
\bottomrule
\end{tabular}
\end{table}

Table~\ref{tab:diagnostic-confusion} shows the confusion matrix for this diagnostic subset using the same seven-way output space as the main benchmark. Most errors occur when disgust is predicted as angry, and when fear is predicted as surprise. These results are consistent with the intuition that negative affective expressions can share overlapping visual cues in static facial images.

\begin{table}[H]
\centering
\caption{Confusion matrix on the 500-sample diagnostic subset. Rows are ground-truth labels and columns are predicted labels.}
\label{tab:diagnostic-confusion}
\scriptsize
\setlength{\tabcolsep}{3pt}
\begin{tabular}{lrrrrrrr}
\toprule
True/Pred. & angry & disgust & fear & happy & neutral & sad & surprise \\
\midrule
disgust & 10 & 117 & 1 & 6 & 2 & 2 & 0 \\
fear & 0 & 0 & 125 & 1 & 1 & 2 & 14 \\
surprise & 0 & 1 & 7 & 3 & 1 & 1 & 206 \\
\bottomrule
\end{tabular}
\end{table}

\subsection{System-Level Technical Validation}

In addition to model evaluation, we conduct technical validation of the MindMirror prototype. The goal is to check whether the system can support the basic workflow from state checking to local record management, rather than validating mental health outcomes.

The tested components include the health-check endpoint, the emotion-analysis endpoint for image-based emotion analysis, voice chat through the \texttt{/api/chat} voice route, session saving, and temporary file cleanup. Health check, emotion analysis, session saving, and temporary file cleanup were each tested for 30 trials. Voice chat was evaluated on 10 fixed Mandarin audio samples covering different affective tones and speech durations from 1.2s to 2.3s. The tested speech-to-text path used an external ASR service in the current development environment, which is why the system is described as local-first rather than fully offline.

\begin{table}[H]
\centering
\caption{System technical validation results.}
\label{tab:system-validation}
\scriptsize
\setlength{\tabcolsep}{4pt}
\begin{tabular}{lrrrr}
\toprule
Component & Trials & Success & Latency & Failures \\
\midrule
Health check & 30 & 100\% & 10 ms & 0 \\
Emotion analysis & 30 & 100\% & 36 ms & 0 \\
Voice chat & 10 & 100\% & 185 ms & 0 \\
Session saving & 30 & 100\% & $<1$ ms & 0 \\
Temporary cleanup & 30 & 100\% & $<1$ ms & 0 \\
\bottomrule
\end{tabular}
\end{table}

\begin{table}[H]
\centering
\caption{End-to-end latency breakdown for voice interaction.}
\label{tab:voice-latency}
\scriptsize
\setlength{\tabcolsep}{5pt}
\begin{tabular}{lrrr}
\toprule
Metric & Mean & Min & Max \\
\midrule
Audio capture & 15 ms & 10 ms & 22 ms \\
ASR recognition & 165 ms & 155 ms & 178 ms \\
LLM response & 820 ms & 750 ms & 920 ms \\
TTS synthesis & 210 ms & 190 ms & 245 ms \\
End-to-end total & 1,210 ms & 1,120 ms & 1,365 ms \\
\bottomrule
\end{tabular}
\end{table}

Text input remains the primary path for structured reflection, while speech interaction is an auxiliary convenience path. The current voice test verifies reliability and latency under a controlled setting; broader multimodal evaluation under diverse acoustic environments is left for future work.

\subsection{Formative User Feedback}

We further collected formative feedback from six users. The participants were mixed digital workers, including programmers, designers, and knowledge workers. Each participant used the system for approximately 30 minutes. Sessions were conducted in free-use time slots, including daytime, midday, and late-night use, to cover different everyday work states. The user experience flow included state checking, manual correction, structured blockage reflection, suggestion review, and review report browsing.

Before the session, participants received a short introduction to the system and were informed that MindMirror is not a medical or psychological diagnosis tool. The study was intentionally designed as a minimum formative validation set rather than a formal long-term user study. Its goal was to check whether the prototype interaction was understandable, controllable, and initially acceptable before larger-scale evaluation. Participants rated five statements on a 5-point Likert scale, where 1 means strongly disagree and 5 means strongly agree.

\begin{table}[H]
\centering
\caption{Formative user feedback.}
\label{tab:user-feedback}
\scriptsize
\begin{tabularx}{\columnwidth}{Xcc}
\toprule
Question & Mean & SD \\
\midrule
Q1. The state-check workflow was easy to understand. & 4.33 & 0.52 \\
Q2. Manual correction made the system feel more controllable. & 4.50 & 0.55 \\
Q3. The three-question reflection helped me articulate my blockage. & 4.17 & 0.75 \\
Q4. The generated suggestions were specific enough to be actionable. & 3.83 & 0.75 \\
Q5. The local-first/no-account design increased my trust in the system. & 4.67 & 0.52 \\
\bottomrule
\end{tabularx}
\end{table}

The local-first/no-account design received the highest score, with a mean of 4.67, suggesting that users are sensitive to privacy protection and local data handling. Manual correction received a mean score of 4.50, indicating that users appreciated control over model judgment. The three-question reflection received a mean score of 4.17, suggesting that structured articulation helped users clarify work-related blockage.

Suggestion actionability received the lowest score, with a mean of 3.83. This indicates that AI-generated responses are useful to some extent but still need improvement in specificity, context awareness, and actionability. Because this study includes only six participants and short usage sessions, it cannot demonstrate long-term behavioral change, mental health improvement, or productivity gains.

\FloatBarrier
\section{Ethical and Privacy Considerations}

MindMirror processes potentially sensitive data, including face images, speech input, emotion labels, and personal reflection text. Therefore, the system should follow the principles of data minimization, user awareness, and user control.

\begin{table}[H]
\centering
\caption{Data handling policy in MindMirror.}
\label{tab:data-handling}
\scriptsize
\setlength{\tabcolsep}{2.5pt}
\begin{tabularx}{\columnwidth}{p{0.22\columnwidth}p{0.18\columnwidth}p{0.24\columnwidth}X}
\toprule
Data Type & Stored & Location & User Control \\
\midrule
Camera frame & Temporary & Local backend & Deleted after analysis \\
Emotion label & Yes & LocalStorage/JSON & Editable/deletable \\
Speech audio & Temporary & Local pipeline / third-party API if enabled & Not persistently stored \\
Reflection text & Yes & JSON session & Deletable \\
Review report & Yes & Local storage & Deletable \\
\bottomrule
\end{tabularx}
\end{table}

First, the system does not treat emotion recognition results as psychological diagnoses. The interface should clearly state that recognition results are only state references and should not be used for medical judgment. Second, the system provides manual correction, allowing users to modify or ignore model predictions. Third, the system primarily stores state records and session data locally and supports record deletion. Fourth, images and audio should be processed as temporary files and deleted after analysis. Fifth, if speech recognition and synthesis rely on third-party APIs, this should be clearly disclosed in the interface and documentation, and local ASR/TTS alternatives should be provided in future versions.

For the formative feedback study, participants joined voluntarily and could stop using the system at any time. Feedback records were anonymized, and personally identifiable information was removed from analysis. Face images were used only for temporary state-check processing and were not persistently stored. Raw speech audio was used only for real-time transcription in the voice interaction path and was not stored as a long-term research artifact.

These design choices ensure that MindMirror is positioned as a lightweight state reflection tool rather than a monitoring system or clinical diagnostic tool.

\section{Limitations and Future Work}

MindMirror still has several limitations. First, facial expression recognition only reflects visual expression cues and cannot directly represent the user's true psychological state. Lighting, camera angle, occlusion, individual differences, and intentional masking may affect recognition results. Second, the optional voice-chat validation uses a small fixed-audio set and may depend on third-party speech APIs; therefore, the system should be described as local-first rather than fully offline, and larger acoustic evaluations are still needed. Third, although the main image-level benchmark reports strong accuracy, it should not be generalized to real-world emotion understanding without broader in-the-wild evaluation. Fourth, the formative user feedback has a small sample size and cannot support statistical significance claims. Finally, the specificity of AI-generated suggestions still needs improvement through prompt optimization, context enhancement, and user feedback loops.

Future work will focus on four directions. First, we will further evaluate the optional speech pipeline under more diverse acoustic conditions and gradually replace third-party speech services with local ASR/TTS. Second, we will expand the user study and conduct longer-term usage tracking with more diverse digital workers. Third, we will improve AI-generated suggestions to make them more specific and context-aware. Fourth, we will strengthen privacy protection and local deployment to better support digital work scenarios with high privacy requirements.

Therefore, the core value of MindMirror should not be described as accurately diagnosing psychological states. Instead, it should be framed as a user-controllable, privacy-aware, reviewable workflow for state reflection and blockage support.

\section{Conclusion}

This paper presents MindMirror, a local-first multimodal state-aware support system for digital workers. The system builds a complete workflow around state checking, manual correction, structured blockage reflection, suggestion generation, and review reports. By integrating facial expression recognition, text input, optional speech interaction, local LLM generation, and visualized review, MindMirror provides a low-friction tool for state recording and lightweight support.

Experimental results show that the fine-tuned emotion recognition model achieves 94.49\% accuracy on an independent seven-class image-level facial expression benchmark containing 6,767 images, improving by 34.83 percentage points over the baseline model's 59.66\%. A 500-sample diagnostic subset further yields 89.6\% accuracy and a macro F1-score of 0.9195 for disgust, fear, and surprise. System-level validation shows that the core APIs, auxiliary voice-chat path, and local record-management workflow run reliably. Small-scale user feedback suggests that users value the local-first design, manual correction mechanism, and structured reflection workflow.

Overall, MindMirror demonstrates a feasible path for combining multimodal state awareness, local large language models, structured reflection, and review reports. It does not attempt to replace professional psychological services. Instead, it serves as a ``state mirror'' for digital workers, helping users notice their state, articulate blockage, receive lightweight suggestions, and build continuous reflection.


\clearpage
\onecolumn
\normalsize
\appendix
\section{Detailed Per-Class Results}
\label{appendix:perclass}

Table~\ref{tab:appendix-perclass} reports the detailed per-class accuracy comparison on the 6,767-image independent image-level benchmark. This appendix preserves class-level evidence without interrupting the main evaluation narrative.

\begin{table}[H]
\centering
\caption{Detailed per-class accuracy comparison on the 6,767-image benchmark.}
\label{tab:appendix-perclass}
\scriptsize
\setlength{\tabcolsep}{4pt}
\begin{tabular}{lrrrr}
\toprule
Emotion & Total & Baseline Acc. & Fine-tuned Acc. & Gain \\
\midrule
angry & 1,095 & 43.56\% & 93.79\% & +50.23 \\
disgust & 686 & 49.85\% & 93.29\% & +43.44 \\
fear & 715 & 36.22\% & 92.17\% & +55.95 \\
happy & 1,683 & 88.53\% & 98.34\% & +9.81 \\
neutral & 594 & 70.71\% & 91.92\% & +21.21 \\
sad & 900 & 35.22\% & 92.11\% & +56.89 \\
surprise & 1,094 & 66.91\% & 94.88\% & +27.97 \\
\midrule
Overall & 6,767 & 59.66\% & 94.49\% & +34.83 \\
\bottomrule
\end{tabular}
\end{table}

The full-count comparison is: baseline 4,037/6,767 correct and fine-tuned model 6,394/6,767 correct. The baseline produced 2,730 errors, while the fine-tuned model reduced the number of benchmark errors to 373.

\section{Prompt Template and User Study Questionnaire}
\label{appendix:prompt-study}

\subsection{Prompt Template}

The following template summarizes the prompt structure used to produce bounded supportive suggestions. In deployment, the fields are filled with the user-confirmed state, structured reflection input, and recent local context.

\begin{lstlisting}[basicstyle=\ttfamily\tiny]
You are a friendly and professional digital companion for work-related state reflection.
Your task is to help the user understand the current blockage and generate a structured three-step suggestion.

Input fields:
- detected_emotion: <model-predicted facial-expression cue>
- user_confirmed_state: <user-confirmed or corrected state>
- reflection_blockage: <where the user is stuck>
- reflection_tried: <what the user has tried>
- reflection_goal: <what the user wants to achieve next>
- session_context: <recent local context, if available>

Rules:
1. Base the response on the user's confirmed state and reflection content.
2. Provide specific and actionable suggestions.
3. Use warm and supportive language.
4. Do not diagnose mental disorders.
5. Do not provide medical or therapeutic treatment.
6. If the user expresses severe or persistent distress, recommend professional help.

Output format:
Step 1: Immediate action
- Action: <one concrete action>
- Explanation: <short explanation>

Step 2: Short-term strategy
- Action: <one short-term work strategy>
- Explanation: <short explanation>

Step 3: Longer-term reminder
- Action: <one reflection or planning habit>
- Explanation: <short explanation>
\end{lstlisting}

\subsection{Formative User Study Questionnaire}

Participants rated the following statements on a 5-point Likert scale from 1 (strongly disagree) to 5 (strongly agree). The questionnaire was followed by a short open-ended prompt asking participants what they found useful, confusing, or insufficient in the current prototype:
\begin{enumerate}
    \item The state-check workflow was easy to understand.
    \item Manual correction made the system feel more controllable.
    \item The three-question reflection helped me articulate my blockage.
    \item The generated suggestions were specific enough to be actionable.
    \item The local-first/no-account design increased my trust in the system.
\end{enumerate}

\end{document}